\documentstyle[aps,prl,preprint,floats,epsfig]{revtex}

\begin{document}
\preprint{\tighten\vbox{\hbox{\hfil CLNS 97/1502}
                        \hbox{\hfil CLEO 97-18}
}}

\footnotesep=0.1cm
\title{Measurement of the Branching Fractions 
of $\Lambda_c^+\to p\overline{K}n(\pi)$}
\author{CLEO Collaboration}
\date{\today}
\maketitle
\tighten

\begin{abstract} 
Using data recorded by the CLEO-II detector at CESR, we 
report new measurements of the branching fractions for the 
decays of the charmed baryon $\Lambda_c^+$ into 
$pK^-\pi^+\pi^0$,
$p\overline{K}^0$,
$p\overline{K}^0\pi^+\pi^-$, and
$p\overline{K}^0\pi^0$,
all measured relative to $pK^-\pi^+$.
The relative branching fractions are $0.67\pm0.04\pm0.11,
0.46\pm0.02\pm0.04,0.52\pm0.04\pm0.05$, and $0.66\pm0.05\pm0.07$
respectively.
\end{abstract}
\newpage
{
\renewcommand{\thefootnote}{\fnsymbol{footnote}}
\begin{center}

\begin{center}
M.~S.~Alam,$^{1}$ S.~B.~Athar,$^{1}$ Z.~Ling,$^{1}$
A.~H.~Mahmood,$^{1}$ H.~Severini,$^{1}$ S.~Timm,$^{1}$
F.~Wappler,$^{1}$
A.~Anastassov,$^{2}$ J.~E.~Duboscq,$^{2}$ D.~Fujino,$^{2,}$
% ...
\end{center}
 
\small
\begin{center}
$^{1}${State University of New York at Albany, Albany, New York 12222}\\
$^{2}${Ohio State University, Columbus, Ohio 43210}\\
% ...
\end{center}

\end{center}

\setcounter{footnote}{0}
}
\newpage

Since the first observation of the lowest lying charmed baryon, the $\Lambda_c^+$, 
there have been many measurements made of its exclusive decay channels. As it is 
difficult to measure the production cross-section of the $\Lambda_c^+$ baryons, 
decay rates are typically presented as branching ratios relative to $\Lambda_c^+
\to pK^-\pi^+$, the most easily observed decay channel. However, fewer than half
of the $\Lambda_c^+$ hadronic decays are presently accounted for. Measurement of these 
modes is of practical as well as theoretical interest. Here, we present measurements of
the branching fractions of $\Lambda_c^+$ into
$pK^-\pi^+\pi^0$,   
$p\overline{K}^0$,
$p\overline{K}^0\pi^+\pi^-$, and $p\overline{K}^0\pi^0$, 
all relative to $pK^-\pi^+$. The last of these is the first measurement of this mode. 
The other modes have been previously
measured but with considerably larger uncertainties than in the present study.

The data presented here 
were taken by the CLEO II detector\cite{KUB} operating at the Cornell 
Electron Storage Ring.
The sample used in this analysis corresponds to
an integrated luminosity of 4.8 $fb^{-1}$ from data
taken on the $\Upsilon(4S)$ 
resonance and in the continuum at energies just above and below 
the $\Upsilon(4S)$.
We detected charged tracks with a cylindrical drift chamber system inside
a solenoidal magnet. Photons were detected using an electromagnetic
calorimeter consisting of 7800 cesium iodide crystals.

Particle 
identification of $p,K^-$, and $\pi^+$ candidates was performed 
using specific ionization measurements in the drift chamber,
and when present, time-of-flight measurements.
For each mass hypothesis, a combined $\chi^2$ probability $P_i$ was formed
($i=\pi,K,p$). Using these probablilities, a normalized 
probability ratio $L_i$ was evaluated, where 
$L_i=P_i/(P_{\pi}+P_K+P_p)$. Well identified protons peaked near 
$P_p=1.0$ while tracks that were identified to not be protons peak 
near $P_p=0.0$. For a track to be used as a proton in this study, we required 
it to have
$L_p>0.8$, which eliminated much of the background,
though with considerable diminution of 
efficiency. For kaons we applied a looser and more efficient
cut of $L_K>0.1$. We have chosen these cuts using a Monte Carlo simulation 
program to maximize the significance of the signals.
The proton identification requirement resulted in an efficiency that is 
strongly momentum dependent, whereas the kaon identification is rather efficient at 
all momenta.
In order to reduce the large combinatoric background, we required
$x_p>0.5$, where $x_p=P_{\Lambda_c}/\sqrt{E^2_{beam} - m^2_{\Lambda_c}}$ 
is the scaled momentum of the $\Lambda_c^+$ candidate. Approximately 60\% of $\Lambda_c^+$
baryons from $c\overline{c}$ continuum events passed this requirement.

The $\overline{K}^0$ candidates were identified in their decay
$K^0_s \to \pi^+\pi^-$, by reconstructing a secondary vertex from the 
intersection of two oppositely charged tracks in the $r-\phi$ plane.
The invariant mass of the $\overline{K^0}$ candidate must lie within 9
${\rm MeV/c^2}$ (around 3 standard deviations ($ \sigma$) 
of its nominal value.

The $\pi^0$ candidates were selected through their decay
$\pi^0 \to \gamma\gamma$ from pairs of well-defined showers in the
CsI calorimeter with a reconstructed invariant mass within $3\sigma$ of the 
$\pi^0$ mass. In order to reduce the combinatorial background,  
each $\gamma$ was required to have an energy of at least 
50 MeV, and the $\pi^0$ was required to have
a momentum of at least $\rm 300\ MeV/c$.

The resulting mass distributions for the 5 modes are shown in Figure 1. Each peak
was fit to the sum of a Gaussian signal distribution with width fixed to
that obtained from CLEO's GEANT based 
Monte Carlo simulation program, and a second
order polynomial background
distribution. The signal widths used and the resulting signal yields are tabulated in
Table 1.

\begin{table}
\begin{tabular}{lcc}
Mode&MC Width (MeV)&Signal\\
\tableline
$pK^-\pi^+$                 &  16  & $10109\pm191$\\
$pK^-\pi^+\pi^0$            &  22   & $2606\pm165$ \\
$p\overline{K}^0$           &  19   & $1025\pm40$ \\
$p\overline{K}^0\pi^+\pi^-$ &  15   & $985\pm65$ \\
$p\overline{K}^0\pi^0$      &  27   & $774\pm52$ \\
\end{tabular}
\caption{The number of $\Lambda_c^+$'s found with $x_p(\Lambda_c)>0.5$}
\end{table}

The efficiency for each $\Lambda_c^+$ mode was calculated using the 
Monte Carlo simulation program\cite{GEANT}. The particle identification efficiency
was checked using real data from $\Lambda \to p\pi^-$ and $D^{*+} \to K^-\pi^+\pi^+$ 
decays
that were identified topologically. The reconstruction efficiency
of the $\Lambda_c^+$ decays has some 
dependence on the resonant substructure of these states. 
In the case of the $pK^-\pi^+$ mode, the Monte-Carlo generator produced a mixture of
non-resonant three-body decay together with 
$\Delta^{++}K^-$ and $p\overline{K}^{*0}$ decays, 
according to their measured branching fractions\cite{PDB}. 
These three types 
of decays had slightly different reconstruction efficiencies, so that including the 
substructure changes the efficiency by $\Delta\epsilon / \epsilon=0.02$
relative to 3-body phase space.
  We have also investigated
the dependence of the reconstruction efficiency of the other modes on possible 
resonant substructure. The poor signal to background ratio did not allow a detailed
measurement of the substructure of these modes.
The efficiency calculation took into account the $\overline{K^0}\to K^0_s$
and $K_s^0 \to \pi^=\pi^-$ branching fractions. 

We have considered many possible sources of  systematic error in the measurement. The
main contributors to the systematic uncertainty came from the following sources: 
1) Uncertainties in the fitting procedures, which were 
estimated by looking at the 
changes in the yields
 using different orders of polynomial background and different 
signal widths (15\% in the case of $pK^-\pi^+\pi^0$, but much smaller for the other modes), 
2) uncertainties due to the unknown mix of resonant substructure 
in the multi-body decays (up to 3\% depending on the mode),
3) uncertainties due to $\pi^0$ finding (5\%), 
$K^0_s$ finding (5\%) and track finding (1\%), and 4) uncertainties in the reconstruction 
efficiency due to the particle identification criteria for protons and kaons (4\%). 
These uncertainties
have been added in quadrature to obtain the total systematic uncertainty for each mode, 
taking into account the fact that many of these tend to cancel in a measurement 
of ratios of branching fractions.  

There are three main types of quark decay diagrams that contribute to $\Lambda_c^+$ decays.
The simplest method is the simple spectator diagram in which the virtual $W^+$ fragments independently of the 
spectator quark. 
The second method involves the
quark daughters of the $W^+$ combining with the remaining quarks. The third method,
W-exchange, involves the $W^+$ combining with the initial $d$ quark. Unfortunately all the
decay modes under investigation here can proceed by more than one of these decay diagrams,
and their decay rates are not amenable to calculation. 

In conclusion, we have measured new branching fractions of the 
$\Lambda_c^+$ into
4 decay modes, measured relative to the normalizing mode $\Lambda_c^+\to pK^-\pi^+$.
The results for three of these modes are in agreement with, and more accurate than, 
previous measurements. We have made 
the first measurement of the decay rate of 
$\Lambda_c^+\to p\overline{K}^0\pi^0$. These measurements help account for the total 
width of the $\Lambda_c^+$ and increase the understanding of charmed baryon decays.  
\begin{table}
\begin{tabular}{llcl}
Mode & Relative Efficiency & $B/B(pK^-\pi^+)$ & Previous Measurements\\
\tableline
$pK^-\pi^+$                 &  1.0     & 1.0& \\

$pK^-\pi^+\pi^0$            &  0.383   & $0.67\pm 0.04\pm 0.11$ &$0.72^{+0.32}_{-0.22}$\cite{BARL}\\
$p\overline{K}^0$           &  0.218   & $0.46\pm 0.02\pm 0.04$&$0.44\pm0.07\pm0.05$\cite{AVE}\\
                            &          &                       &$0.55\pm0.17\pm0.14$\cite{ANJ}\\
                            &          &                       &$0.62\pm0.15\pm0.03$\cite{ALB}\\
$p\overline{K}^0\pi^+\pi^-$ &  0.187   & $0.52\pm 0.04\pm 0.05$&$0.43\pm0.12\pm0.04$\cite{AVE}\\
                            &          &                       &$0.98\pm0.36\pm0.08$\cite{BARL}\\
$p\overline{K}^0\pi^0$      &  0.115   & $0.66\pm 0.05\pm 0.07$&\\
\end{tabular}
\caption{The measured relative branching fractions}
\end{table}

\begin{figure}
\noindent
\epsfig{bbllx=30pt,bblly=130pt,bburx=280pt,bbury=620pt,file=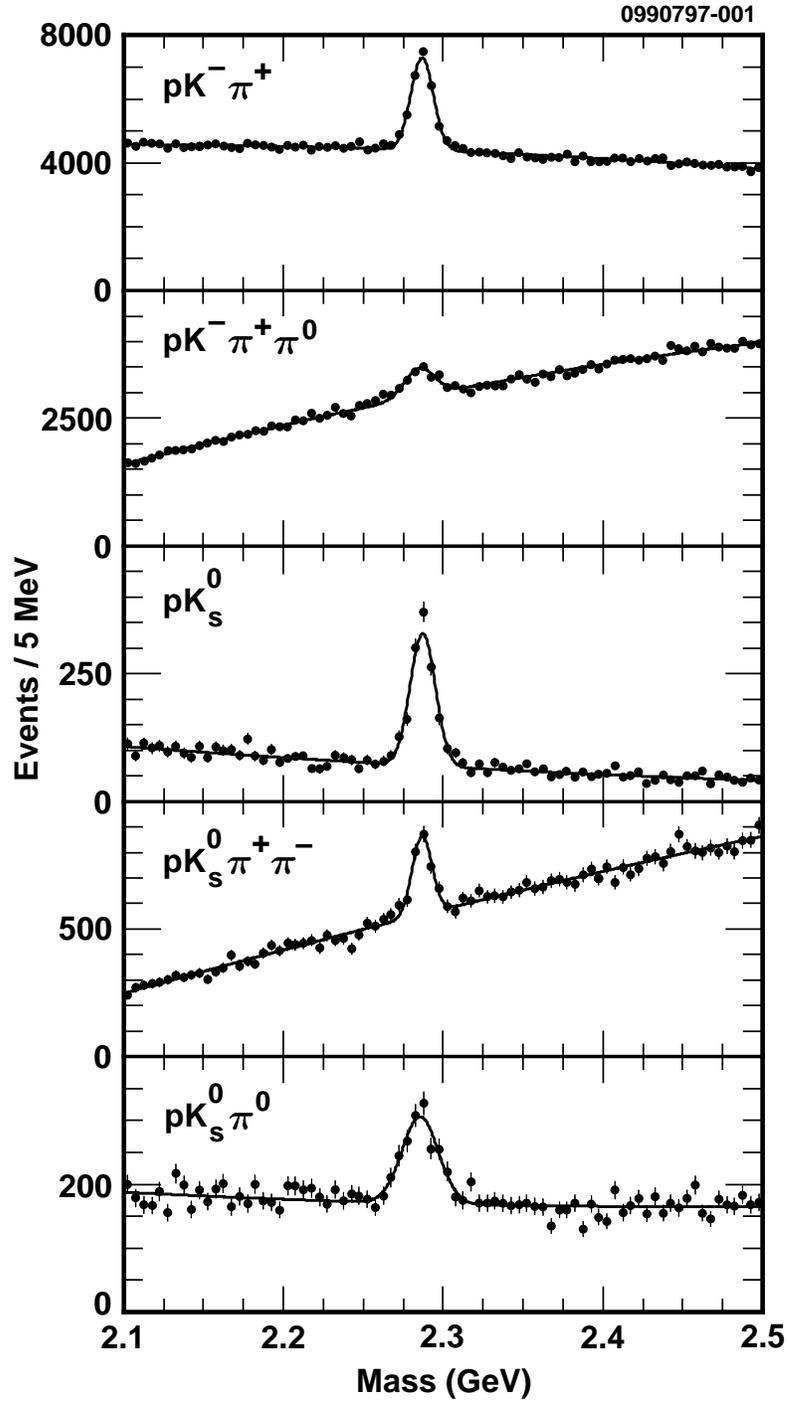,width=3in}
\caption{Invariant mass plots for the 5 different decay modes
of the $\Lambda_c^+$}
\end{figure}

\end{document}